\documentclass[journal]{IEEEtranTIE}
\IEEEoverridecommandlockouts
\usepackage{cite}
\usepackage{amsmath,amssymb,amsfonts}
\usepackage{algorithmic}
\usepackage[ruled,vlined]{algorithm2e}
\usepackage{graphicx}
\usepackage{textcomp}
\usepackage{xcolor}
\usepackage{array}
\def\BibTeX{{\rm B\kern-.05em{\sc i\kern-.025em b}\kern-.08em
    T\kern-.1667em\lower.7ex\hbox{E}\kern-.125emX}}
\usepackage{stfloats}
\usepackage{url}
\usepackage{caption}

\usepackage{booktabs}

\UseRawInputEncoding
\usepackage{listings}
\definecolor{codegreen}{rgb}{0,0.6,0}
\definecolor{codegray}{rgb}{0.5,0.5,0.5}
\definecolor{codepurple}{rgb}{0.58,0,0.82}
\definecolor{backcolour}{rgb}{0.95,0.95,0.92}

\lstdefinestyle{mystyle}{
    backgroundcolor=\color{backcolour},   
    commentstyle=\color{codegreen},
    keywordstyle=\color{magenta},
    numberstyle=\tiny\color{codegray},
    stringstyle=\color{codepurple},
    basicstyle=\ttfamily\footnotesize,
    breakatwhitespace=false,         
    breaklines=true,                 
    captionpos=b,                    
    keepspaces=true,                 
    numbers=left,                    
    numbersep=5pt,                  
    showspaces=false,                
    showstringspaces=false,
    showtabs=false,                  
    tabsize=2
}

\usepackage{orcidlink}

\usepackage{tikz}
\usetikzlibrary{shapes.geometric, arrows}

\tikzstyle{startstop} = [rectangle, rounded corners, minimum width=3cm, minimum height=1cm,text centered, draw=black, fill=red!30]
\tikzstyle{process} = [rectangle, minimum width=3cm, minimum height=1cm, text centered, draw=black, fill=orange!30]
\tikzstyle{arrow} = [thick,->,>=stealth]

\usepackage{tikz}
\usetikzlibrary{positioning, shapes.geometric, arrows.meta}

\tikzstyle{input} = [rectangle, rounded corners, minimum width=3cm, minimum height=1cm, text centered, draw=black, fill=blue!10]
\tikzstyle{block} = [rectangle, rounded corners, minimum width=3cm, minimum height=1cm, text centered, draw=black, fill=green!10]
\tikzstyle{arrow} = [thick,->,>=stealth]

\makeatletter
\newcommand{\linebreakand}{%
  \end{@IEEEauthorhalign}
  \hfill\mbox{}\par
  \mbox{}\hfill\begin{@IEEEauthorhalign}
}
\makeatother

\graphicspath{{./Figures}}

\begin{document}




	\title{Deep Reinforcement Learning for Optimizing Inverter Control: Fixed and Adaptive Gain Tuning Strategies for Power System Stability}


	\author{{Shuvangkar Chandra Das, Tuyen Vu, Deepak Ramasubramanian, Evangelos Farantatos, Jianhua Zhang, Thomas Ortmeyer\\
	\textit{}}
		
		\thanks{Corresponding author: dassc@clarkson.edu}}

	



\maketitle

\begin{abstract}
	This paper presents novel methods for tuning inverter controller gains using deep reinforcement learning (DRL). A Simulink-developed inverter model is converted into a dynamic link library (DLL) and integrated with a Python-based RL environment, leveraging the multi-core deployment and accelerated computing to significantly reduce RL training time. A neural network-based mechanism is developed to transform the cascaded PI controller into an actor network, allowing optimized gain tuning by an RL agent to mitigate scenarios such as subsynchronous oscillations (SSO) and initial transients. Two distinct tuning approaches are demonstrated: a fixed gain strategy, where controller gains are represented as RL policy (actor network) weights, and an adaptive gain strategy, where gains are dynamically generated as RL policy (actor network) outputs. A comparative analysis of these methods is provided, showcasing their effectiveness in stabilizing the transient performance of grid-forming and grid-following converters and deployment challenges in hardware. Experimental results are presented, demonstrating the enhanced robustness and practical applicability of the RL-tuned controller gains in real-world systems.
	
\end{abstract}
 

\section*{Nomenclature}
\begin{tabular}{@{}p{0.15\linewidth}p{0.8\linewidth}@{}}
	\toprule
	\textbf{Symbol} & \textbf{Description} \\
	\midrule
	$\theta$ & Policy (ANN Actor) parameters\\
	$\phi$ & Value function (ANN Critic) parameters \\
	$\mathcal{B}$ & Rollout buffer for storing trajectory data \\
	$\pi_\theta$ & Policy function parameterized by $\theta$ \\
	$a_t$ & Action taken at time step $t$ \\
	$s_t$ & State observations at time step $t$ \\
	$r_t$ & Reward received at time step $t$ \\
	$L^{CLIP}$ & Clipped surrogate objective (policy loss) \\
	$L^{VF}$ & Value function loss \\
	$S[\pi_\theta]$ & Entropy of the policy $\pi_\theta$ \\
	$\text{KL}$ & Kullback-Leibler divergence \\
	$\hat{A}_t$ & Advantage estimate at time $t$ \\
	$\alpha$ & Learning rate \\
	$\epsilon$ & PPO clipping range\\
	\bottomrule
\end{tabular}

\section{Introduction}

	Although artificial intelligence has revolutionized the tech industry, its adoption in the power systems domain has been relatively slow. In particular,  AI technique, such as deep reinforcement learning (DRL), has seen limited exploration for power system control, especially in addressing stability issues in power inverters. Ref. \cite{massaoudiNavigatingLandscapeDeep2023} presents a comprehensive review of published research in this area from 2005 to 2023. The data shows a sharp rise in RL-based power system stability papers starting in 2019, peaking at 80 publications in 2022. Growing interest highlights the potential of RL to tackle complex challenges in power system stability, though further research is necessary to broaden its applications in this field.

	Ref. \cite{xiongDeepReinforcementLearning2022, wooDSTATCOMDqAxis2021, schenkeControllerDesignElectrical2020, saadatmandAdaptiveCriticDesignbased2021, matlabTrainTD3Agent} explored using RL agents to replace traditional PI controllers in various inverter-based applications. However, this approach presents several key challenges: (1) The DRL model generates control signals directly, which is difficult to implement in hardware due to the high computational demand of the neural network based policy, making it impractical for embedded systems. (2) Additionally, the policy may respond unpredictably to scenarios, not encountered during training, potentially leading to system instability under untested conditions. In contrast, ref. \cite{baltasGridFormingPowerConverters2020, sheInverterPQControl2024,guanDesignReinforcementLearning2020} used RL agents to tune PI controllers by adjusting adaptive gains. This method enhances the approach of replacing the PI controller with a DRL agent by retaining the PI controller and enabling adaptive tuning. However, it still faces hardware deployment limitations, as the RL policy requires a fully connected neural network, which is computationally intensive and challenging to implement in real-time systems.



	One of the prominent challenges of EMT simulation is the requirement for very small time steps to accurately solve the network dynamics. For instance, Simulink, a widely used software for EMT simulation, requires approximately 20,000 computational steps per second, which demands significant computational resources to ensure accurate results. On the other hand, the computation methods for RL value functions are divided into Monte Carlo (MC) and Temporal Difference (TD) methods \cite{chenReviewApplicationsReinforcement2024}. MC method estimate value function by averaging the returns from complete episode, providing higher accuracy due to their reliance on the full trajectories of data. 
	However, it is not always feasible to obtain complete sample data for every episode, especially in the field of power electronic converters, where the training process may be interrupted by unsafe operating conditions. Also, it is desirable to update the value function and control policy in real-time during the training process. From the existing literature, it can be observed that currently in the field of power electronics, there is a greater emphasis on the application of RL algorithms based on TD methods \cite{chenReviewApplicationsReinforcement2024}. Therefore, the RL algorithm needs to run 20,000 computation per second together with the inverter based power system environment.  If the model runs for 5s for each episode, it requires 100,000 computation for each episode. Our experiments indicate that the RL agent requires a significant number of training episodes, ranging from 50 to 2,000, to achieve reasonable convergence.


	\begin{figure*}[!t] 
		\centering 
		\includegraphics[width=1.0\textwidth]{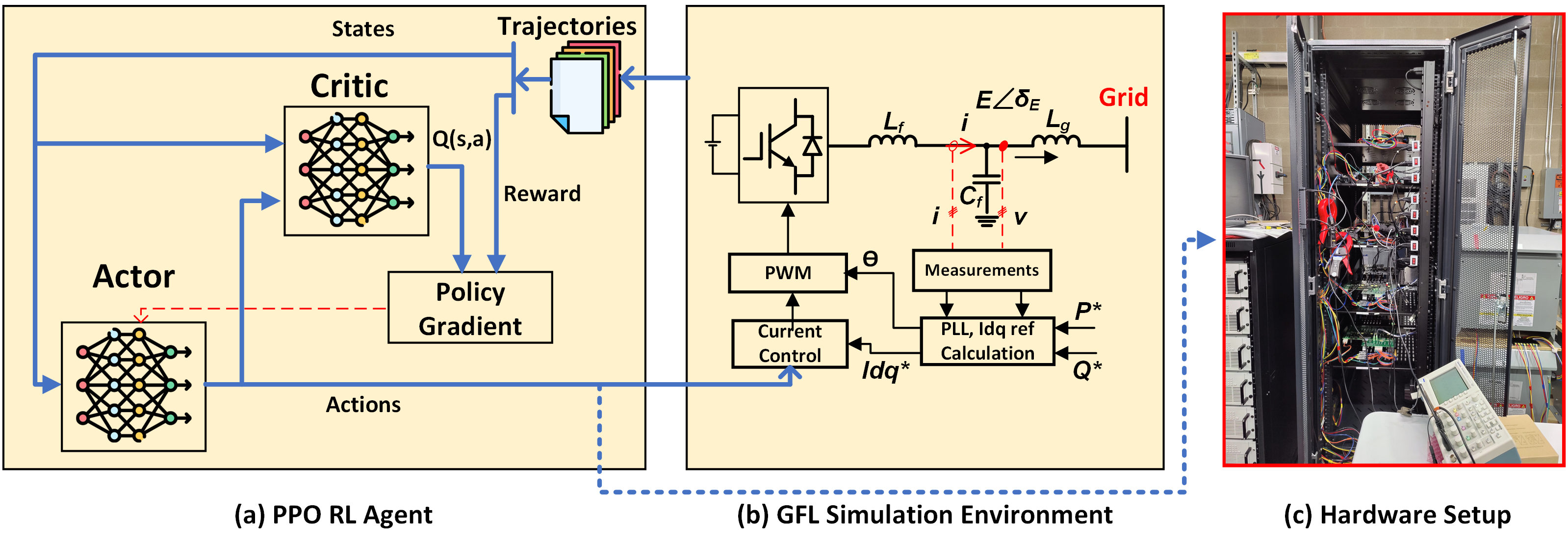} 
		\caption{GFL inverter RL agent training pipeline where(a) represents the PPO Agent, (b) GFL inverter environment setup with the grid (GFM, load and transformer), (c) Trained gains / RL policy is transported into hardware setup} 
		\label{fig:DRL-GFM-environment} 
	\end{figure*}
	
	\begin{table*}[!b]
		\centering
		\caption{State-of-art DRL training environment for IBR controller tuning}
		\label{tab:EMT-RL-Training-comparison}
		\begin{tabular}{p{3cm} p{3.5cm} p{4cm} p{4cm} p{1.5cm}}
			\hline
			\textbf{Environment Name} & \textbf{Process} & \textbf{Advantages} & \textbf{Disadvantages} & \textbf{Ref} \\
			\hline
			Native MATLAB Environment & EMT modeling in Simulink and RL agent in MATLAB & Low build time & Higher training time & \cite{alejandro-sanjinesAdaptivePIController2023}\\
			\hline
			Python Environment & Model inverter and RL agent in Python& High flexibility & Difficult building EMT model of IBR in Python & \cite{balakrishnaGymelectricmotorGEMPython2021}\\
			\hline
			MATLAB + Python Hybrid Environment & IBR model in Simulink and RL Agent in Python. Use communication to transfer data & Rich Python RL library, easy EMT modeling in Simulink & High communication time and high training time & \cite{sheInverterPQControl2024} \\
			\hline
			Native Python Environment (DLL-based) & Model the IBR in Simulink, convert into multiple DLLs, Build the RL agent and environment in Python & Lower training time. Able to parallel compute the EMT model and GPU acceleration of RL agent training. High flexibility & Initial challenge of interfacing IBR DLL into Python environment. Developed a Python library to overcome this challenge & \cite{schaeferPythonBasedReinforcementLearning2024} \\
			\hline
		\end{tabular}
		
	\end{table*}

Various EMT simulation platforms, such as MATLAB, offer integration capabilities for developing reinforcement learning (RL) agents, allowing easy integration with simulation environments \cite{matlabReinforcementLearningToolbox}. While these platforms can significantly reduce development time, they may result in longer training durations compared to other programming environments. For instance, previous studies developed electric motor EMT models and RL agents in Python-based environments, demonstrating computational efficiency, though these approaches often require more time to develop the EMT models \cite{balakrishnaGymelectricmotorGEMPython2021, traueReinforcementLearningEnvironment2022}. Another approach in the literature involves developing the RL agent in a programming environment like Python while using an EMT simulation platform such as MATLAB for modeling the system \cite{sheInverterPQControl2024}. However, this method introduces higher communication latency, leading to longer training times. A comparison of these approaches is presented in Table \ref{tab:EMT-RL-Training-comparison}.

In this research, we addressed these key challenges of IBR controller tuning using DRL method. While DRL-based automatic controller tuning holds promise for optimizing gains in non-linear systems like IBRs, two major obstacles persist: (1) the high training time for such models in EMT simulation software (MATLAB environment) and (2) the impracticality of replacing traditional PI controllers with RL agents or deploying RL agents for adaptive gain tuning due to computational limitations in microcontrollers. To overcome these challenges, we propose a hybrid approach that leverages the strengths of both Simulink and Python. The EMT model is developed in Simulink and converted into a DLL, enabling integration with a Python-based RL environment using the PyTorch library. This approach allows us to utilize the best of both worlds; Simulink's advanced modeling capabilities and Python's computational power and rich RL libraries offering a more efficient and practical solution for controller tuning in IBR-based power system. Finally, contributions of this research are as follows:

\begin{itemize}
	
    \item A novel neural network-based mechanism is developed to convert the inverter cascaded PI controller into an actor network, enabling gain optimization by an RL agent under scenarios such as transients, subsynchronous oscillation (SSO) and faults. The tuned gains can be directly deployed into inverters, ensuring robust performance and seamless integration with existing systems.
    
	\item Two approaches are developed for tuning inverter control gains: a fixed gain method, where controller gains are embedded as weights of actor network, and an adaptive gain method, where gains are generated dynamically as actor network outputs. A comparison is provided, highlighting the effectiveness of both strategies in stabilizing the transient performance of grid-forming and grid-following converters.
	

	\item A pipeline is proposed for IBR controller tuning, where the inverter model developed in an EMT simulation platform (e.g., Simulink) is converted into a DLL and integrated into a programming environment (e.g., Python) for reinforcement learning (RL). This approach utilizes multi-core deployment and accelerated computing to optimize the model, significantly reducing training time. To the best of our knowledge, this is the first application of a DLL-based IBR controller tuning mechanism employing deep reinforcement learning (DRL), setting a new precedent in the field.

    \item Presented experimental results showcasing the improved performance of RL-tuned controller gains, highlighting their practical application and impact.
\end{itemize}

\begin{figure*}[!t] 
	\centering 
	\includegraphics[width=1.0\textwidth]{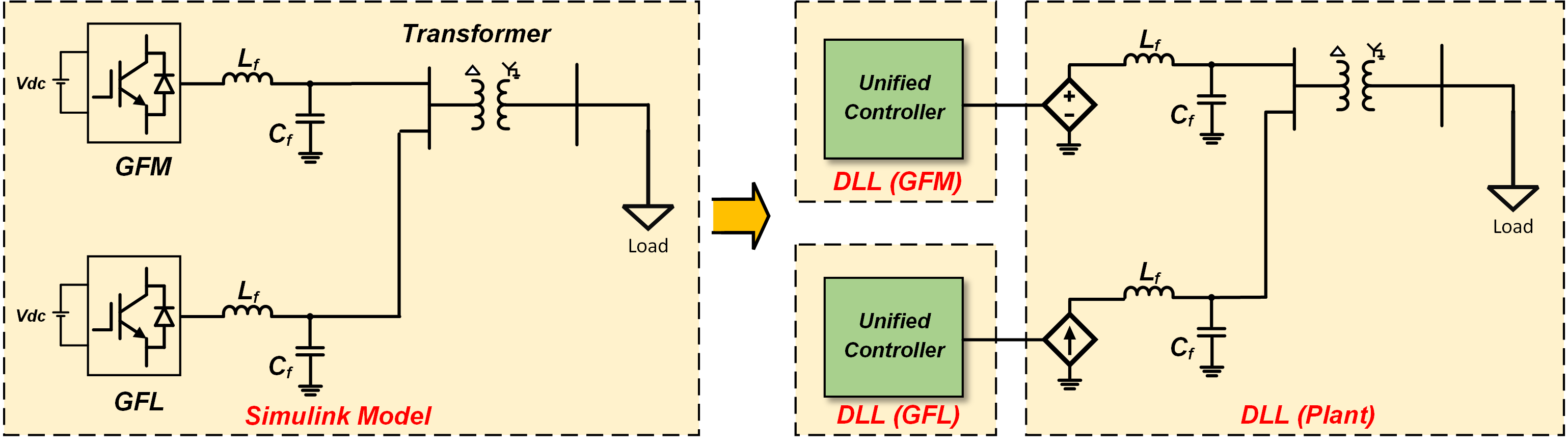} 
	\caption{Simulink to DLL generation for parallel computing of the RL environment} 
	\label{fig:inverter-to-dll} 
\end{figure*}


The organization of this paper is as follows: section \ref{method} outlines the mechanism for building the inverter environment and developing the RL agent, section \ref{results} presents the performance of the DRL-based controller tuning for the GFL inverter and compares the two methods explored in this research, section \ref{discussion} covers the applicability, advantages, and disadvantages of both methods, while section \ref{conclusion} offers the final remarks.

\section{DRL Pipeline for IBR Controller Tuning}
\label{method}
The IBR optimal controller gain tuning problem is 
formulated as a Markov decision
process (MDP) \cite{suttonReinforcementLearningIntroduction1998b} depicted in the Fig. \ref{fig:DRL-GFM-environment}. The system state represents key variables such as dq-axis current, error signal, the integration of the error signal etc. The DRL agent generates actions, which corresponds to generating dq-axis voltage reference from the current controller. These actions are based on the current state of the system. The environment, which consists of the inverter and the power grid, transitions to a new state in response to these actions. A reward function is designed to reflect system performance, rewarding actions that improve stability and penalizing those that lead to oscillations or instability. This setup, with states, actions, reward, and transitions, ensures that the control problem can be efficiently solved using RL within the MDP framework.

\subsection{IBR Power System Environment Formulation}

\begin{figure}[!t] 
	\centering 
	\includegraphics[width=0.5\textwidth]{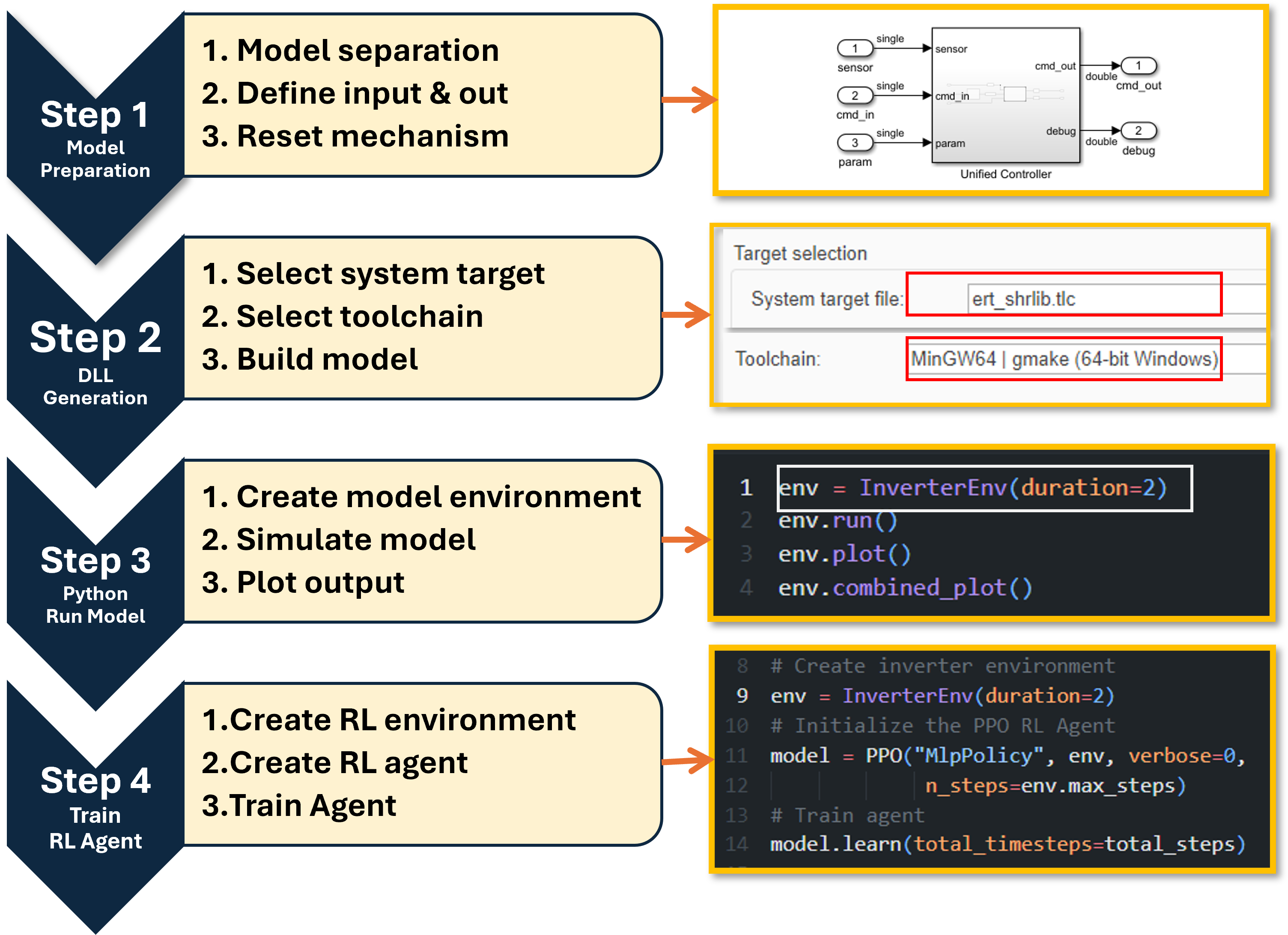} 
	\caption{Flowchart of 4 steps process of Simulink Model DRL training pipeline in Python environment} 
	\label{fig:RL-Training-Process} 
\end{figure}

The IBR-based power system environment explored in this research consists of two inverters, as shown in Fig. \ref{fig:inverter-to-dll}. One grid-forming (GFM) and one grid-following (GFL) inverter supply power to a single resistive load. The system is divided into three zones: the grid-forming controller, the grid-following controller, and the electric network, which includes the voltage source, current source, filters, and load. This subdivision aims to deploy different parts of the grid across multiple cores to accelerate training while also enhancing modularity and flexibility. The process of building the environment for RL agent training involves four steps, as depicted in Fig. \ref{fig:RL-Training-Process}.

\subsubsection{\textbf{Model Preparation}}


The first step in preparing the Simulink model for DLL generation is dividing the system into distinct subsystems, as shown in Fig. \ref{fig:inverter-to-dll}. This modular approach enables more efficient deployment across multiple CPU cores, with each subsystem generating its own DLL. By distributing the computation workload, this method accelerates the training process compared to converting the entire model into a single DLL. The separation ensures efficient data exchange between subsystems, while maintaining the circuit within one subsystem.

The second part of model preparation focuses on defining the model's inputs and outputs, as these are crucial for establishing the RL environment. Depending on the project objectives, the input and output signals must be clearly specified. In the case of the inverter environment, the model includes three inputs: sensor feedback, user command input, and control parameters. The outputs consist of command signals for controlling the power switches and debug signals for monitoring the inverter's performance.



The third step involves integrating a reset signal into the model. During training, the environment must reset at the end of each episode. While EMT software such as  Simulink provides automatic reset functionality, this feature must be manually incorporated when exporting the model to a Python environment. For example, at the end of each episode, integrators need to be reset, and components such as capacitors and inductors must be discharged. These reset functions are triggered by an input signal. In this test setup, a rising edge signal resets the controller. For the plant, resetting is achieved by setting the DC input voltage of the inverter to zero and running the plant model for a few cycles to discharge energy storage elements.


\subsubsection{\textbf{DLL Generation}}

The literature offers various approaches for converting EMT model (Simulink) into DLLs, but many of these methods are complex, involving multiple steps and manual intervention. For example, ref. \cite{pfeifferSimulinkModelDLL} outlines a process that requires manual coding, increasing both complexity and the risk of errors. Similarly, ref. \cite{bonetSimulinkPython} describes a multi-step approach where the Simulink model is first converted into C code, followed by the use of the GCC compiler to generate the DLL. This method adds extra stages, making the process more time-consuming and prone to mistakes.


In contrast, the method presented in this research simplifies the process significantly by converting the EMT model into a DLL using a simulation platform's code generation capabilities (e.g., MATLAB/Simulink  \cite{matlabEmbeddedCoderGenerate} or PSCAD) and an appropriate compiler (e.g., MinGW64). This approach requires minimal manual intervention: once the compiler and target file are set, the DLL can be built directly from the simulation model using the platform's code generation tools. This streamlined method saves time and reduces the potential for errors, making it efficient and practical for real-world applications, as motivated by recent work in \cite{schaeferPythonBasedReinforcementLearning2024}.

\subsubsection{\textbf{RL Environment Creation}}
The third step of the proposed method involves wrapping the DLL using Python and building the RL training environment. This process includes developing a Python script to interact with the DLL, creating an appropriate interface for input and output signals. The RL training environment is then set up using the Python Gymnasium module \cite{foundationGymnasiumDocumentation}, a fork of the original Gym library developed by OpenAI \cite{openaiGymDocumentation}. Additionally, several graph plotting methods were created to visualize the data during training, aiding in system debugging.

\subsubsection{\textbf{RL Agent Formation}}
The steps involve building the environment instance, creating the RL agent, and training the model. Python offers many choices for this purpose. In this research, PyTorch-based Stable Baselines 3 \cite{raffinStableBaselines3ReliableReinforcement2021, raffinStableBaselines3DocsReliable2021} was used to develop the RL agent. Two types of DRL agents are formulated to obtain the fixed and adaptive gains for the GFL current controller, as explained in subsections \ref{fixed-gain-rl-model} and \ref{adaptive-gain-rl-model}.

\begin{figure*}[!th] 
	\centering 
	\includegraphics[width=1.0\textwidth]{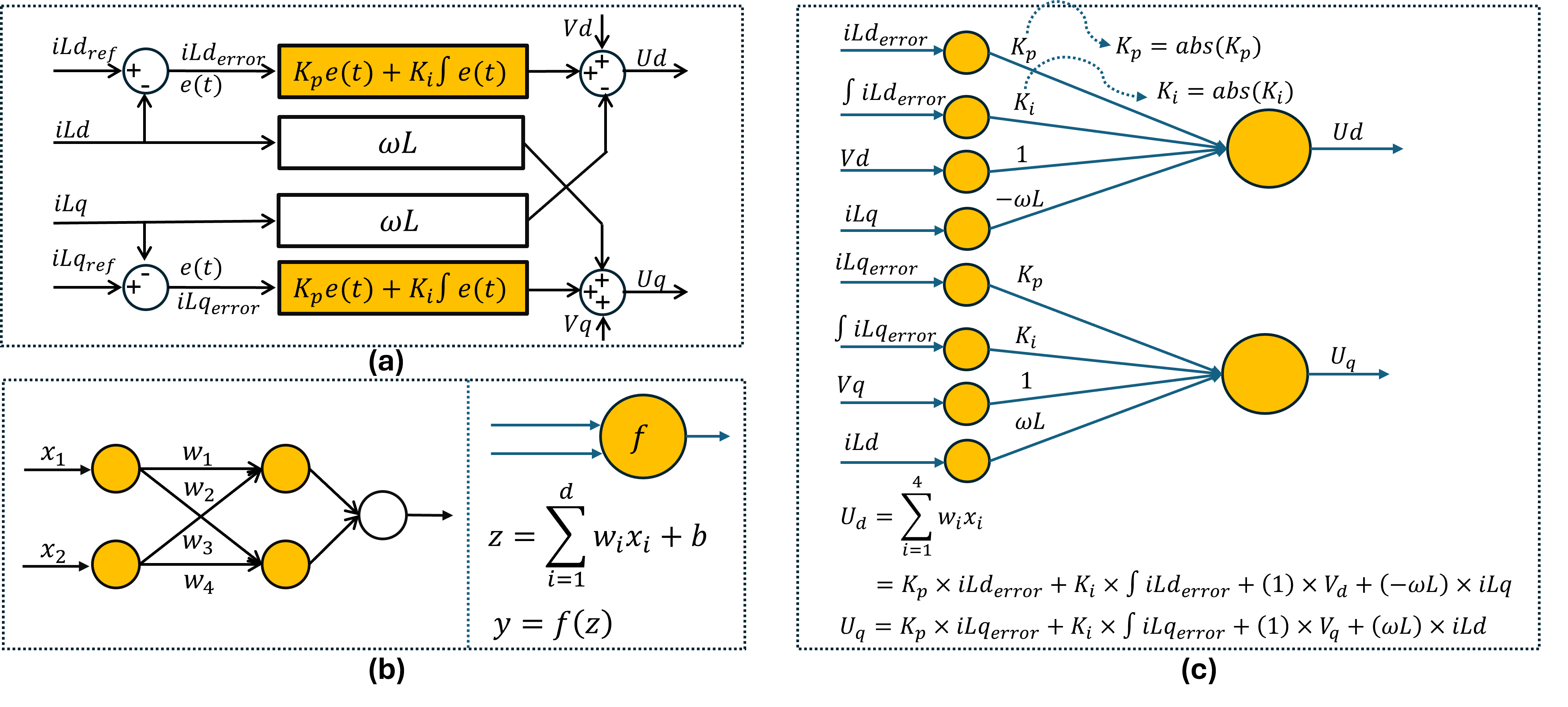} 
	\caption{Process of converting cascaded current controller into neural network for actor. (a) shows a conventional current controller (b) shows a simple fully connected neural network. each node takes input from all previous layers inputs calculate z, then passed through activation function calculates final output (c) Represetns current controller as neural network. The goal is set the PI gains, $K_p$ and $K_i$ as the neural network weights.} 
	\label{fig:Current-PI-to-NN} 
\end{figure*}


\subsection{DRL Algorithm Implementation}

Two types of DRL agents are formulated to obtain fixed and adaptive gains for the GFL current controller. Both of the DRL models utilize the Proximal Policy Optimization (PPO) algorithm, with a shared training mechanism. The primary differences lie in their hyperparameters and actor-critic designs. This subsection focuses on the common elements of the training mechanism. First, a brief overview of the PPO algorithm, which is used to address this control problem, is provided, along with the implementation details

\subsubsection{\textbf{PPO Algorithm}}

\begin{algorithm}[!b]
	\caption{PPO Training strategy}
	\label{algo:PPO}
	\SetAlgoLined
	\nl Initialize policy parameters $\theta$\;
	\nl Initialize value function parameters $\phi$\;
	\nl Initialize rollout buffer $\mathcal{B}$\;
	\nl \For{iteration = 1, 2, ..., total\_iterations}{
		\nl \For{step = 1, 2, ..., n\_steps}{
			\nl $a_t \sim \pi_\theta(a_t|s_t)$\;
			\nl Execute action $a_t$ and observe $r_t$, $s_{t+1}$\;
			\nl Store $(s_t, a_t, r_t, V_\phi(s_t), \log\pi_\theta(a_t|s_t))$ in $\mathcal{B}$\;
		}
		\nl Compute returns and advantages for all steps in $\mathcal{B}$\;
		\nl \For{epoch = 1, 2, ..., n\_epochs}{
			\nl \For{minibatch in $\mathcal{B}$}{
				\nl Compute policy (actor) loss $L^{CLIP}$\;
				\nl Compute value function (critic) loss $L^{VF}$\;
				\nl Compute entropy bonus $S[\pi_\theta]$\;
				\nl $L = -L^{CLIP} + c_1 L^{VF} - c_2 S[\pi_\theta]$\;
				\nl Update $\theta$ and $\phi$ by minimizing $L$\;
			}
		}
		\nl Clear rollout buffer $\mathcal{B}$\;
		\nl \If{KL divergence $>$ target\_kl}{
			\nl Early stopping\;
			\nl Break\;
		}
	}
	\nl \Return Optimized policy parameters $\theta$
\end{algorithm}

Proximal Policy Optimization (PPO), developed by John Schulman in 2017 \cite{schulmanProximalPolicyOptimization2017a}, has become the default reinforcement learning algorithm at OpenAI due to its simplicity and robust performance \cite{opanaiProximalPolicyOptimization}. PPO is widely regarded for its general applicability and improved sample efficiency, demonstrated empirically in various tasks. The key objective function in the PPO algorithm is the clipped surrogate objective function, $L^{C L I P}$, which constrains policy updates within a small range by applying a clipping mechanism.

\begin{equation}
	L^{C L I P}(\theta)=\hat{\mathbb{E}}_t\left[\min \left(r_t(\theta) \hat{A}_t, \operatorname{clip}\left(r_t(\theta), 1-\epsilon, 1+\epsilon\right) \hat{A}_t\right)\right]
\end{equation}

\noindent
where epsilon is a hyperparameter, say, $\epsilon=0.2$. The motivation for this objective is as follows. The first term inside the equation is $L^{C P I}$
	
	\begin{equation}
		L^{C P I}(\theta)=\hat{\mathbb{E}}_t\left[\frac{\pi_\theta\left(a_t \mid s_t\right)}{\pi_{\theta_{\text {old }}}\left(a_t \mid s_t\right)} \hat{A}_t\right]=\hat{\mathbb{E}}_t\left[r_t(\theta) \hat{A}_t\right]
	\end{equation}

	\begin{equation}
		r_t(\theta)=\frac{\pi_\theta\left(a_t \mid s_t\right)}{\pi_{\theta_\theta \text { old }}\left(a_t \mid s_t\right)}, 
	\end{equation}
 
\noindent
Here, $r_t(\theta)$ denotes the probability ratio 
If $r_t(\theta)>1$, the action $a_t$ at state $s_t$ is more likely in the current policy than the old policy. If $r_t(\theta)$ is between 0 and 1 , the action is less likely for the current policy than for the old one. So this probability ratio is an easy way to estimate the divergence between old and current policy.

\noindent
The second term, $\operatorname{clip}\left(r_t(\theta), 1-\epsilon, 1+\epsilon\right) \hat{A}_t$, modifies the surrogate objective by clipping the probability ratio, which removes the incentive for moving $r_\iota$ outside of the interval $[1-\epsilon, 1+\epsilon]$. Then, we take the minimum of the clipped and non-clipped objective, so the final objective is a lower bound (pessimistic bound) of the unclipped objective.

\subsubsection{\textbf{PPO Implementation}}

	Algorithm \ref{algo:PPO} presents the implementation of the PPO algorithm. The process iterates through environment interactions and model updates. In each iteration, the algorithm collects data by sampling actions from the current policy $\pi_\theta$ for n steps, storing state-action-reward tuples $(s_t, a_t, r_t, \pi_\theta(a_t|s_t))$ in a buffer trajectory $\mathcal{B}$ depicted in Fig. \ref{fig:DRL-GFM-environment}. It then computes advantages for each step and trains the agent over multiple epochs. During training, the algorithm calculates policy and value function losses, and updates the actor and critic network parameters by minimizing a combined loss function. This approach allows for stable and efficient policy optimization in reinforcement learning tasks.
	
\subsection {Fixed Gain DRL Model Formulation}
\label{fixed-gain-rl-model}
\subsubsection{\textbf{Modeling Cascaded PI Controller into Actor Network}}


	A PI controller can be modeled as a single-layer neural network with error and integral of error as inputs. In \cite{matlabTunePIControllera}, this neural network representation is used to control a water tank system. To prevent negative weights during optimization, the weights are constrained by applying an absolute value function: \( Y = |w_i| \times X \). The control signal is then computed as:



	\begin{equation}
		u = \begin{bmatrix} \int e \, dt & e \end{bmatrix} 
			\begin{bmatrix} K_i \\ K_p \end{bmatrix}
	\label{eq:matrix_PI_controller}
	\end{equation}

	However, representing a more complex control system, such as the cascaded PI controller used in GFM and GFL voltage and current control, as a neural network was unclear. To address this challenge, this research proposes a mechanism of converting a cascaded PI controller into a neural network representation. Fig. \ref{fig:Current-PI-to-NN} illustrates this process using three sub-figures. A simple single-layer fully connected neural network is represented in Fig. \ref{fig:Current-PI-to-NN}(b). Each node takes inputs from all previous nodes, multiplies them by corresponding weights, and sums them with a bias term. This operation can be expressed by the following equation:

	\begin{equation}
	z = \sum_{i=1}^d w_i x_i + b
	\label{eq:nn-eqn}
	\end{equation}


	where \(x_i\) represents the input, \(w_i\) is the weight of each connection, and \(b\) is the bias. After computing \(z\), it is passed through an activation function \(f(z)\) to calculate the final output of the node. The activation function allows the network to learn non-linear behaviors. Without an activation function, the network would only be capable of learning linear behaviors. This property is utilized in designing the neural network for the PI controller.

	Finally, the \ref {fig:Current-PI-to-NN}(c) represents the GFL current controller as neural network. The goal is to set the PI gains, $K_p$ and $K_i$ as the neural network weights. Ignoring the bias, the input of the each node is represented as:
	\begin{equation}
		\begin{aligned}
		U_d & =\sum_{i=1}^4 w_i x_i \\
		&= w_1\times x_1 + w_2\times x_2 + w_3 \times x_3 + w_4 \times x_4 \\ 
		& =K_p \times i L d_{\text {error }}+K_i \times \int i L d_{\text {error }}+(1) \times V_d+ \\ 
		& (-\omega L) \times i L q 
		\end{aligned}
	\end{equation}

	\noindent
	Similarly, $U_q$ can be represented as neural network as follows: 
	\begin{equation}
		\begin{aligned}
		U_q & =K_p \times i L q_{\text {error }}+K_i \times \int i L q_{\text {error }}+(1) \times V_q + \\
		&(\omega L) \times i L d
		\end{aligned}
	\end{equation}

	\noindent
	The neural network is designed in such a way that only the connection weighs are represented as the PI gains $K_p$ and $K_i$. The rest of the connections weights are represented as a constant values. Since the goal is to learn the linear behaviours of PI controller, hence no activation function is utilized in this network architecture. The abs() function is applied to the weights. It is ensuring that the weights used in the computation are always positive. This is done as negative gains would lead to unstable behavior.

\subsubsection{\textbf{Critic Network for Fixed-gain-DRL-model}}


The critic network in the PPO algorithm is designed to estimate the value function, which is crucial for advantage estimation. It consists of a feedforward neural network with three main layers. The input layer accepts an 8-dimensional state representation, which is processed through two hidden layers, each comprising 64 neurons and followed by ReLU activation functions. These hidden layers allow the network to capture complex relationships within the state space. The final output layer reduces the 64-dimensional representation to a single scalar value, representing the estimated value of the current system state.

The hyperparameters used in this fixed-gain DRL model, such as learning rate, batch size, and discount factor, are outlined in Table \ref{tab:ppo_hyperparameters}. These hyperparameters were carefully chosen to ensure stable training of the PPO algorithm. Additionally, the detailed architecture of the policy network, including both the actor and critic components, is shown in Table \ref{tab:policy_structure}. This structure, with its specific layer sizes and initial values for the control gains \(K_p\) and \(K_i\), ensures efficient representation of the control policies during training.

\begin{table}[htbp]
	\centering
	\caption{PPO hyperparameters of fixed-gain-DRL-model}
	\begin{tabular}{p{5cm} p{3cm}}
	\hline
	\textbf{Hyperparameter} & \textbf{Value} \\
	\hline
	Learning Rate & 0.0003 \\
	Number of Steps & 48000 \\
	Batch Size & 64 \\
	Number of Epochs & 10 \\
	Discount Factor ($\gamma$) & 0.99 \\
	GAE Lambda & 0.95 \\
	Entropy Coefficient & 0.0 \\
	Value Function Coefficient & 0.5 \\
	Max Gradient Norm & 0.5 \\
	\hline
	\end{tabular}%

	\label{tab:ppo_hyperparameters}
\end{table}
	
\begin{table}[htbp]
	\centering
	
	\caption{Policy Structure of Fixed-gain-DRL-model }
	\begin{tabular}{p{3cm} p{5cm}}
	\hline
	\textbf{Component} & \textbf{Structure} \\
	\hline
	Actor & PIActorNetwork (Fig. \ref{fig:Current-PI-to-NN}c) \\
	& $K_p = 1$ (Initial value) \\
	& $K_i = 5$ (Initial value) \\
	\hline
	Critic & Linear(8, 64) \\
	& ReLU() \\
	& Linear(64, 64) \\
	& ReLU() \\
	& Linear(64, 1) \\
	\hline
	\end{tabular}%
	
	\label{tab:policy_structure}
\end{table}



\subsubsection{\textbf{Observations of Fixed-gain-DRL-model}}

In the fixed-gain tuning model, the RL agent utilizes a predefined set of observations to generate actions that tune the GFL inverter's current controller. Fig.  \ref{fig:Current-PI-to-NN}(c) shows the GFL current controller observations, which serve as inputs to the neural network used by the RL agent. The observation set for the fixed-gain model is defined as:
\begin{equation}
    \begin{split}
        \mathcal{O} = \{iLd_{\text{error}}, \int iLd_{\text{error}}, iLq_{\text{error}}, \\
        \int iLq_{\text{error}}, V_d, V_q, iL_d, iL_q\}
    \end{split}
\end{equation}


Table \ref{tab:fixed-gain-observation} summarizes the minimum and maximum values for each observation used during RL agent training. In real-world applications, inverters are halted if these limits are violated, ensuring operational safety. Therefore, during training, the RL agent is tasked with maintaining the system within these limits.



\begin{table}[!h]
	\centering
	\caption{Fixed gain model observation set with minimum and maximum values}
	\begin{tabular}{p{1cm} p{1cm} p{1cm} p{4cm}}
	\hline
	\textbf{Name} & \textbf{Min Value (pu)} & \textbf{Max Value (pu)} & \textbf{Description} \\
	\hline
	$iLd_{\text{error}}$ & $-2$ & $2$ & $iLd_{\text{error}} = iLd_{ref} - iLd$ \\
	$\int iLd_{\text{error}}$ & $-\infty$ & $\infty$ & Integral of the direct axis inductance current error. \\
	$iLq_{\text{error}}$ & $-2$ & $2$ & $iLq_{\text{error}} = iLq_{ref} - iLq$ \\
	$\int iLq_{\text{error}}$ & $-\infty$ & $\infty$ & Integral of the quadrature axis inductance current error. \\
	$V_d$ & $-2$ & $2$ & Direct axis voltage component. \\
	$V_q$ & $-2$ & $2$ & Quadrature axis voltage component. \\
	$iL_d$ & $-2$ & $2$ & Direct axis current. \\
	$iL_q$ & $-2$ & $2$ & Quadrature axis current. \\
	\hline
	\end{tabular}
	
	\label{tab:fixed-gain-observation}
\end{table}


\subsubsection{\textbf{Action Space of Fixed-gain-DRL-model}}
In the fixed gain model, the RL agent replaces the current control block of the GFL inverter. Since the agent mimics the functionality of the current controller, it generates two key actions to drive the inverter: \( U_d \) (direct-axis output voltage) and \( U_q \) (quadrature-axis output voltage). These voltages, \( U_{dq} \), control the inverter switches, as shown in Fig. \ref{fig:Current-PI-to-NN}(c). The \( U_{dq} \) voltages are transformed into \( U_{abc} \) voltages to drive the average inverter model. Table \ref{table:fixed-gain-action} provides a summary of the fixed gain action space, including the minimum and maximum values.

\begin{table}[tb!]
	\centering
	\caption{Observation with Minimum and Maximum Values}
	\label{table:fixed-gain-action}
	\begin{tabular}{p{1cm} p{1cm} p{1cm} p{4cm}}
	\hline
	\textbf{Name} & \textbf{Min Value} & \textbf{Max Value} & \textbf{Description} \\
	\hline
	$U_d$ & $-2.0$ & $2.0$ & direct-axis output voltage \\
	$U_q$ & $-2.0$ & $2.0$ & quadrature-axis output voltage\\
	\hline
	\end{tabular}
	
\end{table}


\subsubsection{\textbf{Reward Function for Fixed-gain-DRL-model}}
Reward measures the immediate effectiveness of taking a particular action. In general, a positive reward to encourage certain agent actions and a negative reward (penalty) to discourage other actions. A common strategy is to provide a small positive reward for each time step that the agent successfully performs the task and a large penalty when the agent fails. Additionally, physical safety constraint terms are incorporated into the reward function to limit overcurrent situations \cite{chenReviewApplicationsReinforcement2024}.

In a discrete-time system with dynamics \( x_{t+1} = A x_t + B u_t \) and initial condition \( x_0 = x^{\text{init}} \), the goal is to choose control inputs \( u_0, u_1, \dots \) that balance two objectives: keeping the system states \( x_0, x_1, \dots \) small for good regulation and minimizing control effort \( u_0, u_1, \dots \). These are often competing goals, as large inputs can reduce state deviations quickly but require more effort.The Linear Quadratic Regulator (LQR) addresses this by minimizing a quadratic cost function:
\[
J(u) = \sum_{\tau=0}^{N-1} \left( x_\tau^T Q x_\tau + u_\tau^T R u_\tau \right) + x_N^T Q_f x_N
\]
\noindent
Here, \( Q \) penalizes state deviations, \( R \) penalizes control effort, and \( Q_f \) is the final state cost. The LQR solution optimally balances these objectives through the control inputs \( u_0, u_1, \dots \).
As the system improves its performance, the LQR cost \( J(U) \) decreases over time. Consequently, the RL reward \( R_t(u) = -J(u) \) increases over time, reflecting better system control and reduced input effort. This approach effectively ties the LQR framework to RL, where maximizing the cumulative reward corresponds to minimizing the traditional LQR cost. QR reward structure encourages an agent to drive observations vector $s$ to zero with minimal action effort. Smooth continuous rewards, such as the QR regulator, are good for fine-tuning parameters and can provide policies similar to optimal controllers (LQR/MPC) \cite{DefineRewardObservation}. 

The reward function is designed to guide the RL agent in generating control signals for the GFL current controller, ensuring accurate current tracking, stable operation, and efficient power production. The reward function \( R_t \) is defined as follows:
\begin{equation}
	\label{eqn:fixed-gain-reward}
	\begin{aligned}
		R_t &= Q_1 \times (iLd_{ref} - iLd)^2 + Q_2 \times (iLq_{ref} - iLq)^2 \\
		    &\quad + Q_3 \times a_1^2 + Q_4 \times a_2^2 \\
			&\quad + Q_5 \times \sum_{i =1}^2 |a_i -LPF(a_i)| + \\
			&\quad + 
		    \begin{cases} 
		    	Q_6 \times |P| & \text{if }  P < 0 \\
		    	0 & \text{otherwise}
		    \end{cases}
	\end{aligned}
\end{equation}


\begin{table}[tb!]
	\centering
	\caption{Fixed-gain-DRL-model reward weights}
	\label{table:fixed-gaing-reward-weights}
	\begin{tabular}{p{2cm} p{1.5cm} | p{2cm} p{1.5cm}}
	\hline
	\textbf{Reward Weight} & \textbf{Value} & \textbf{Reward Weight} & \textbf{Value} \\
	\hline
	\( Q_1 \) & -1 & \( Q_4 \) & -0.1 \\
	\( Q_2 \) & -1 & \( Q_5 \) & -5 \\
	\( Q_3 \) & -0.1 &  \( Q_6 \) & -1 \\
	\hline
	\end{tabular}
	
\end{table}


\noindent
In this function, the terms represent different aspects of the system's performance. The current tracking error terms, \( Q_1 \times (iLd_{ref} - iLd)^2 \) and \( Q_2 \times (iLq_{ref} - iLq)^2 \), penalize deviations between the reference and actual current in the direct and quadrature axes, ensuring that the RL agent minimizes current tracking errors. This is essential for maintaining the desired operation of the inverter. 

The terms \( Q_3 \times a_1^2 \) and \( Q_4 \times a_2^2 \) penalize the magnitude of the control actions \( a_1 \) (direct-axis voltage \( U_d \)) and \( a_2 \) (quadrature-axis voltage \( U_q \)), preventing the RL agent from generating excessively large control signals. This keeps the system stable and efficient by avoiding extreme voltage commands that could destabilize the inverter.

To discourage oscillatory behavior, the term \( Q_5 \times \sum_{i=1}^2 |a_i - LPF(a_i)| \) penalizes the difference between the current action and its low-pass filtered (LPF) version, which smooths the control signal and reduces high-frequency oscillations. By penalizing rapid fluctuations in the control signals, the RL agent is encouraged to produce steady, gradual control actions, which is important for maintaining inverter stability.

Finally, the real power penalty \( Q_6 \times |P| \) is applied when the inverter consumes real power (i.e., when \( P < 0 \)). Since the goal of the GFL inverter is to supply real power, this penalty discourages the agent from consuming power. If the inverter is supplying real power (i.e., \( P \geq 0 \)), no penalty is applied.

The reward weights \( Q_1, Q_2, Q_3, Q_4, Q_5, Q_6 \), as presented in Table \ref{table:fixed-gaing-reward-weights}, balance the importance of each term. Larger penalties for current tracking and oscillation penalties ensure that the RL agent focuses on maintaining accurate current control and stable operation. The smaller penalties on the action magnitudes and power consumption further fine-tune the agent’s behavior to ensure that it generates efficient and practical control signals.


\subsection{Adaptive-gain-DRL-model Formulation}
\label{adaptive-gain-rl-model}
\subsubsection{\textbf{Actor-critic Network}}
The actor-critic architecture employed in adaptive gain model utilizes a similar initial structure for both the actor and critic components, followed by separate output layers. The network consists of two fully connected layers, each with 64 neurons, using hyperbolic tangent (tanh) activation functions. This structure allows for efficient feature extraction from the 4-dimensional input state. The actor network, responsible for policy decisions, culminates in a final linear layer with 2 outputs, corresponding to the adaptive gain parameters. The critic network, tasked with value estimation, concludes with a single-output linear layer. This architecture enables the model to simultaneously learn a policy for adaptive gain control and estimate the value of states, facilitating effective learning in the complex inverter control environment. The use of tanh activations in the hidden layers provides non-linearity while maintaining gradient flow, which is particularly beneficial for control tasks where smooth output transitions are desirable. Table \ref{tab:adaptive_ppo_hyperparameters} shows the hyperparameters used for the PPO algorithm in our adaptive gain model. The structure of our policy, including both the actor and critic networks, is detailed in Table \ref{tab:adaptive_policy_structure}.

\begin{table}[!tb]
	\caption{Adaptive-gain-model PPO Hyperparameters}
	\begin{tabular}{p{5cm} p{3cm}}
	\hline
	\textbf{Hyperparameter} & \textbf{Value} \\
	\hline
	Learning Rate & Dynamic \\
	Number of Steps & 1024 \\
	Batch Size & 64 \\
	Number of Epochs & 10 \\
	Discount Factor ($\gamma$) & 0.99 \\
	GAE Lambda & 0.95 \\
	Clip Range & Dynamic \\
	Entropy Coefficient & 0.0 \\
	Value Function Coefficient & 0.5 \\
	Max Gradient Norm & 0.5 \\
	\hline
	\end{tabular}
	
	\label{tab:adaptive_ppo_hyperparameters}
\end{table}

\begin{table}[!tb]
	\caption{Adaptive-gain-model Policy Structure}
	\begin{tabular}{p{5cm} p{3cm}}
	\hline
	\textbf{Component} & \textbf{Structure} \\
	\hline
	Actor & Linear(4, 64) \\
	& Tanh() \\
	& Linear(64, 64) \\
	& Tanh() \\
	& Linear(64, 2) \\
	\hline
	Critic & Linear(4, 64) \\
	& Tanh() \\
	& Linear(64, 64) \\
	& Tanh() \\
	& Linear(64, 1) \\
	\hline
	\end{tabular}
	
	\label{tab:adaptive_policy_structure}
\end{table}

\subsubsection{\textbf{Adaptive-gain-DRL-model Observations}}

In the adaptive gain tuning model, the RL agent dynamically generates adaptive gains to adjust the PI controller of the GFL inverter's current controller. The goal in this model is to minimize both real and reactive power errors, with the observation set consisting of variables related to power generation and errors. The observation set for the adaptive gain model is defined as:
\begin{equation}
    \begin{split}
        \mathcal{O} = \{P_{\text{gen}}, Q_{\text{gen}}, P_{\text{error}},
        Q_{\text{error}}\}
    \end{split}
\end{equation}

\begin{table}[!bt]
    \centering
	\caption{Adaptive gain model observation set with minimum and maximum values}
    \begin{tabular}{p{1cm} p{1cm} p{1cm} p{4cm}}
    \hline
    \textbf{Name} & \textbf{Min Value} & \textbf{Max Value} & \textbf{Description} \\
    \hline
    $P_{\text{gen}}$ & $-2$ & $2$ & GFL real power generation\\
    $Q_{\text{gen}}$ & $-2$ & $2$ & GFL reactive power generation \\
    $P_{\text{error}}$ & $-2$ & $2$ & GFL real power tracking error \\
    $Q_{\text{error}}$ & $-2$ & $2$ & GFL reactive power tracking error\\
    \hline
    \end{tabular}
    \
    \label{tab:adaptive-gain-observation}
\end{table}

Table \ref{tab:adaptive-gain-observation} outlines the minimum and maximum values of the observations. As with the fixed-gain model, these limits are critical for ensuring the safe operation of the GFL inverter, and violations during training result in episode termination and penalties.

\subsubsection{\textbf{Adaptive-gain-DRL-model Action Space}}
In the adaptive gain model, the RL agent generates the gains for the GFL current controller. The agent produces two adaptive gains: \( K_p \) (proportional gain) and \( K_i \) (integral gain). These gains adjust the PI controller to optimize the inverter's response to real and reactive power requirements. Table \ref{table:adaptive-gain-action} provides a summary of the adaptive gain action space, along with the associated minimum and maximum values.

\begin{table}[tb!]
	\centering
	\caption{Adaptive Gain Model Action Space with Minimum and Maximum Values}
	\label{table:adaptive-gain-action}
	\begin{tabular}{p{1cm} p{1cm} p{1cm} p{4.2cm}}
	\hline
	\textbf{Action} & \textbf{Min Value} & \textbf{Max Value} & \textbf{Description} \\
	\hline
	$K_p$ & $0$ & $20$ & Current Controller Proportional gain \\
	$K_i$ & $0$ & $100$ & Current Controller Integral gain \\
	\hline
	\end{tabular}
\end{table}


\subsubsection{\textbf{Reward Function for Adaptive Gain Model}}

The reward function for the adaptive gain model is designed to guide the RL agent in tuning the proportional (\(K_p\)) and integral (\(K_i\)) gains of the PI controller to minimize real and reactive power tracking errors. It is defined as: 
\begin{equation}
	\label{eqn:adaptive-gain-reward}
	\begin{aligned}
		R_t &= Q_1 \times (P_{ref} - P_{gen})^2 + Q_2 \times (Q_{ref} - Q_{gen})^2 \\ 
		    &\quad + Q_3 \times a_1^2 + Q_4 \times a_2^2 
	\end{aligned}
\end{equation}

In this equation, the terms \( Q_1 \times (P_{ref} - P_{gen})^2 \) and \( Q_2 \times (Q_{ref} - Q_{gen})^2 \) penalize deviations in real and reactive power, ensuring accurate power generation, with weights \( Q_1 = 10 \) and \( Q_2 = 5 \), indicating higher importance for real power. The terms \( Q_3 \times a_1^2 \) and \( Q_4 \times a_2^2 \) penalize large control actions for the proportional (\(a_1= K_p \)) and integral (\(a_2=K_i \)) gains to prevent instability, with small weights \( Q_3 = 0.1 \) and \( Q_4 = 0.1 \), allowing for fine-tuning while maintaining control signal regularity.

Unlike the fixed-gain model, the adaptive-gain model utilizes real and reactive power errors. While using the d and q-axis currents could yield similar results, the real and reactive power errors are preferred to keep the model simpler.

\section{Results}
\label{results}
In this research, experiments are  conducted to determine two types of optimal gains by training two deep reinforcement learning (RL) agents. Since the Simulink model was ported into the Python environment, we first discuss the validation of the model's performance in Python environment. Then, fixed-gain model results are presented, subsequently adaptive gain model result will be discussed.

\subsection{Benchmarking IBR Simulation}


	Since the IBR power system is modeled in Simulink and subsequently converted to Python through DLL conversion, it is crucial to conduct a benchmark test to ensure the Python environment accurately represents the Simulink model. The IBR model comprises one GFM and one GFL inverter supplying a resistive load. At \( t = 0.5s \), the GFL inverter is paralleled with the GFM inverter with a reference power of \( P_{ref} = 0.5 \, pu \). The initial gains from Table \ref{tab:RL-agent-fixed-gains}, specifically \( K_p = 1 \) and \( K_i = 5 \), are applied to the GFL current controller to generate the comparison results.

\begin{figure}[!t] 
		\centering 
		\includegraphics[width=0.48\textwidth]{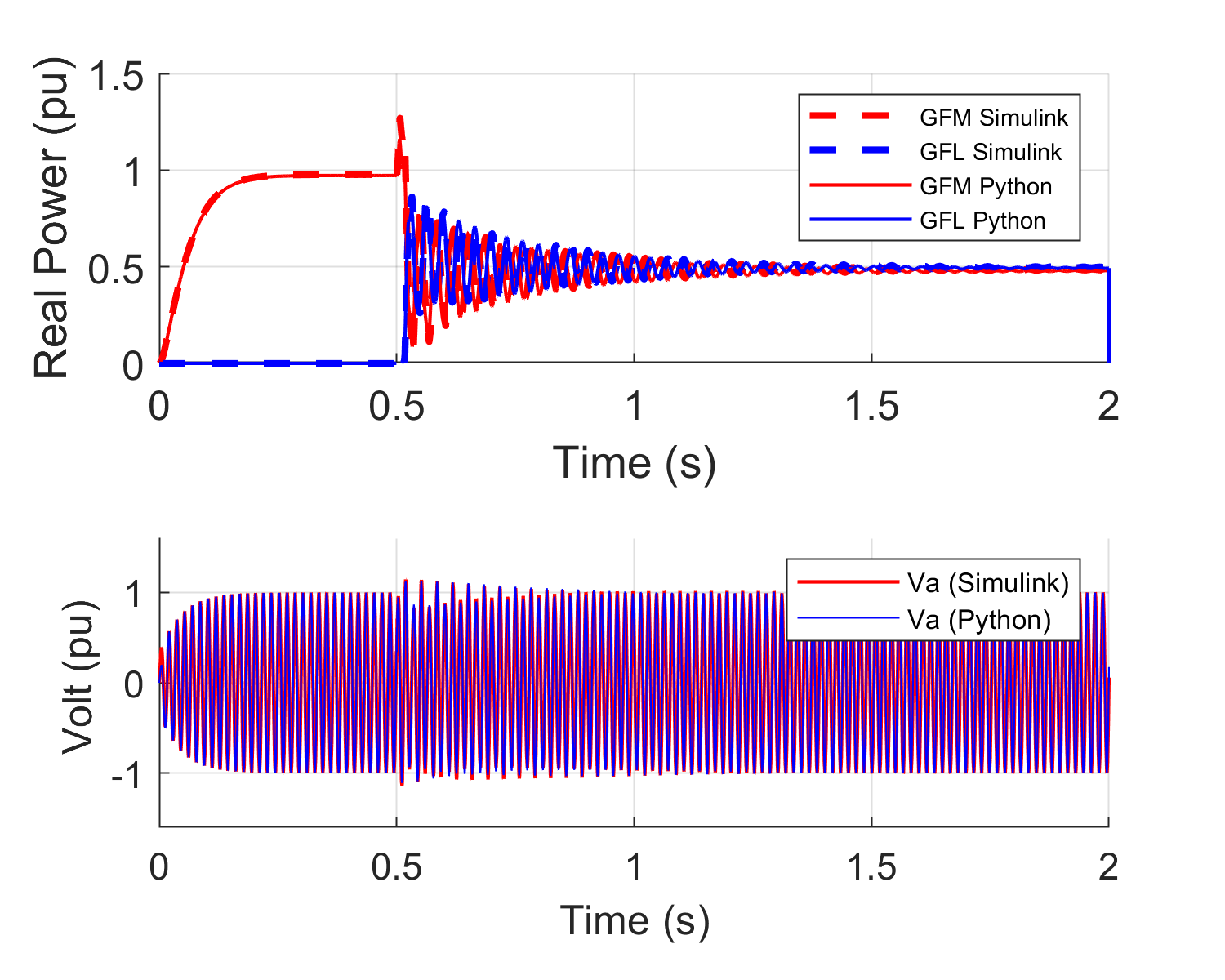} 
		\caption{Comparison result of the IBR in Simulink and Python Environment} 
		\label{fig:sim_python} 
	\end{figure}

	Fig. \ref{fig:sim_python} illustrates the benchmark result obtained using both Simulink and Python environments. The dotted line represents the Simulink simulation results, while the solid line corresponds to the Python simulation results for the same model keeping all other conditions same. The comparison demonstrates that the Python simulation yields results identical to those from Simulink, with the added advantages of enabling reinforcement learning (RL) agent training within the Python environment and the ability to run the simulation model on multiple cores, thereby accelerating both simulation and training times.

	The real power figures confirm that the Python simulation precisely replicates the transient responses observed in the Simulink simulation. Additionally, the phase A voltage signals from both simulations match exactly. Moreover, it is found that the Python simulation runs faster than the Simulink simulation for the same model. This performance gain is attributed to the use of pre-compiled DLLs, which reduce the compilation overhead inherent in Simulink. Furthermore, logging data into scopes in Simulink is computationally expensive, further contributing to the efficiency advantage of the Python implementation.

\subsection{Fixed-gain-DRL-model Simulation Performance}






\begin{figure*}[!ht]
	\centering
	\begin{minipage}{0.45\textwidth}
		\centering 
		\includegraphics[width=1\textwidth]{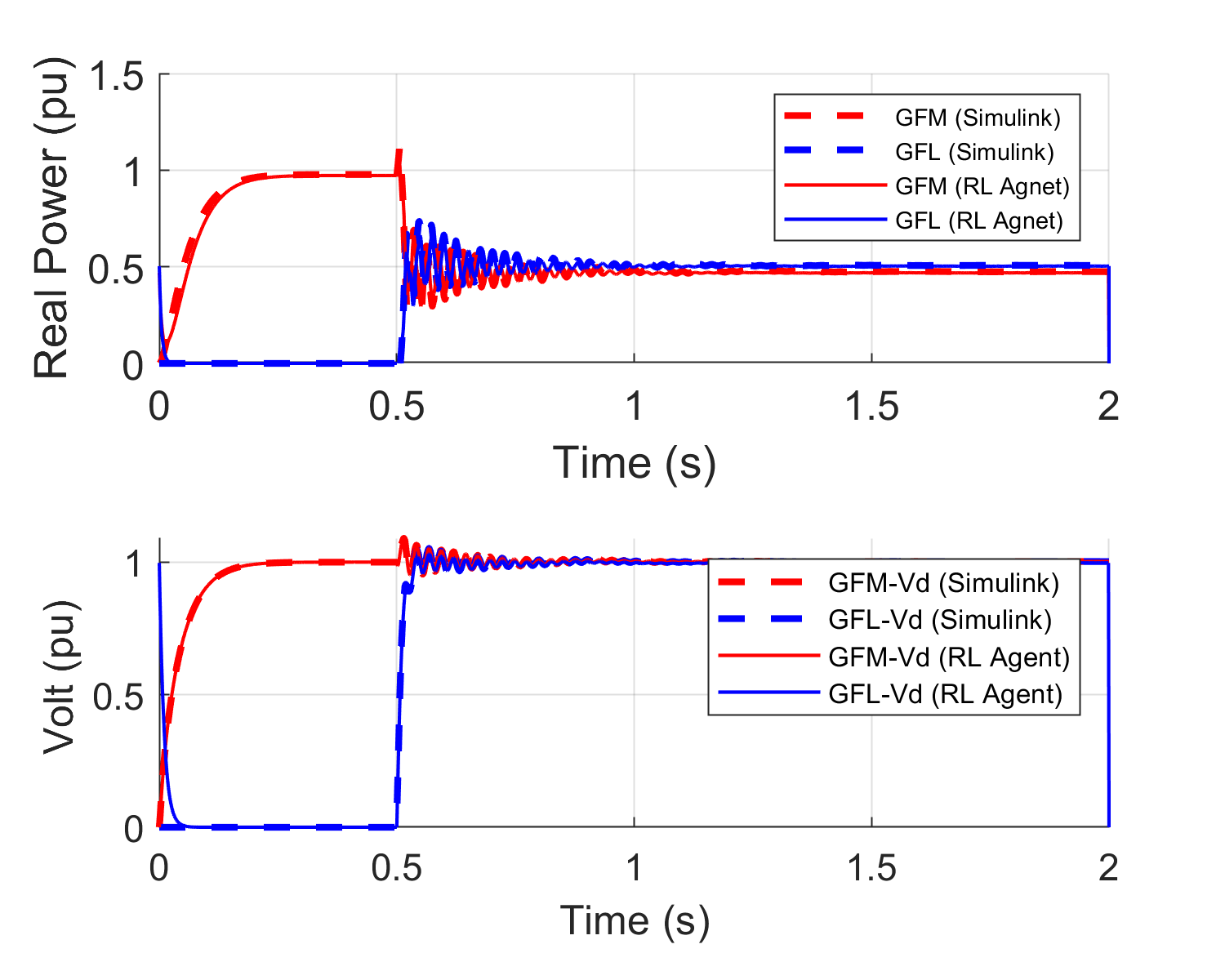} 
		\caption{Performance of GFM and GFL inverters from the fixed-gain-DRL-model} 
		\label{fig:fixed-gain-gfl-gfm-performance} 
	\end{minipage}\hfill
	\begin{minipage}{0.45\textwidth}
		\centering
		\includegraphics[width=1.0\textwidth]{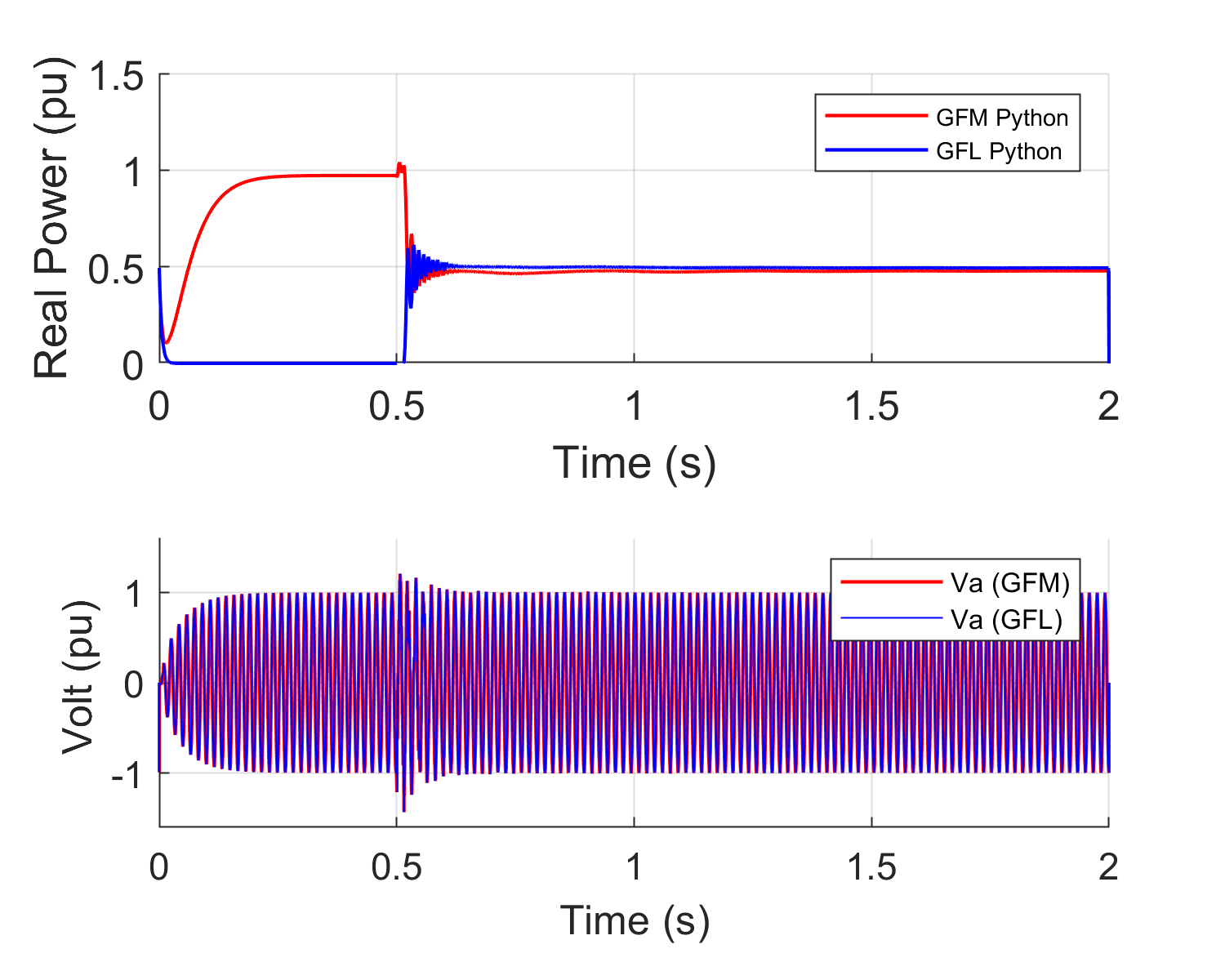}
		\caption{Performance of GFM and GFL inverters from the adaptive-gain-DRL-mode}
		\label{fig:adaptive-gain-performance-python-single-run}
	\end{minipage}
\end{figure*}	

\begin{figure*}[!ht]
	\centering
	\begin{minipage}{0.48\textwidth}
		\centering 
		\includegraphics[width=1.0\textwidth]{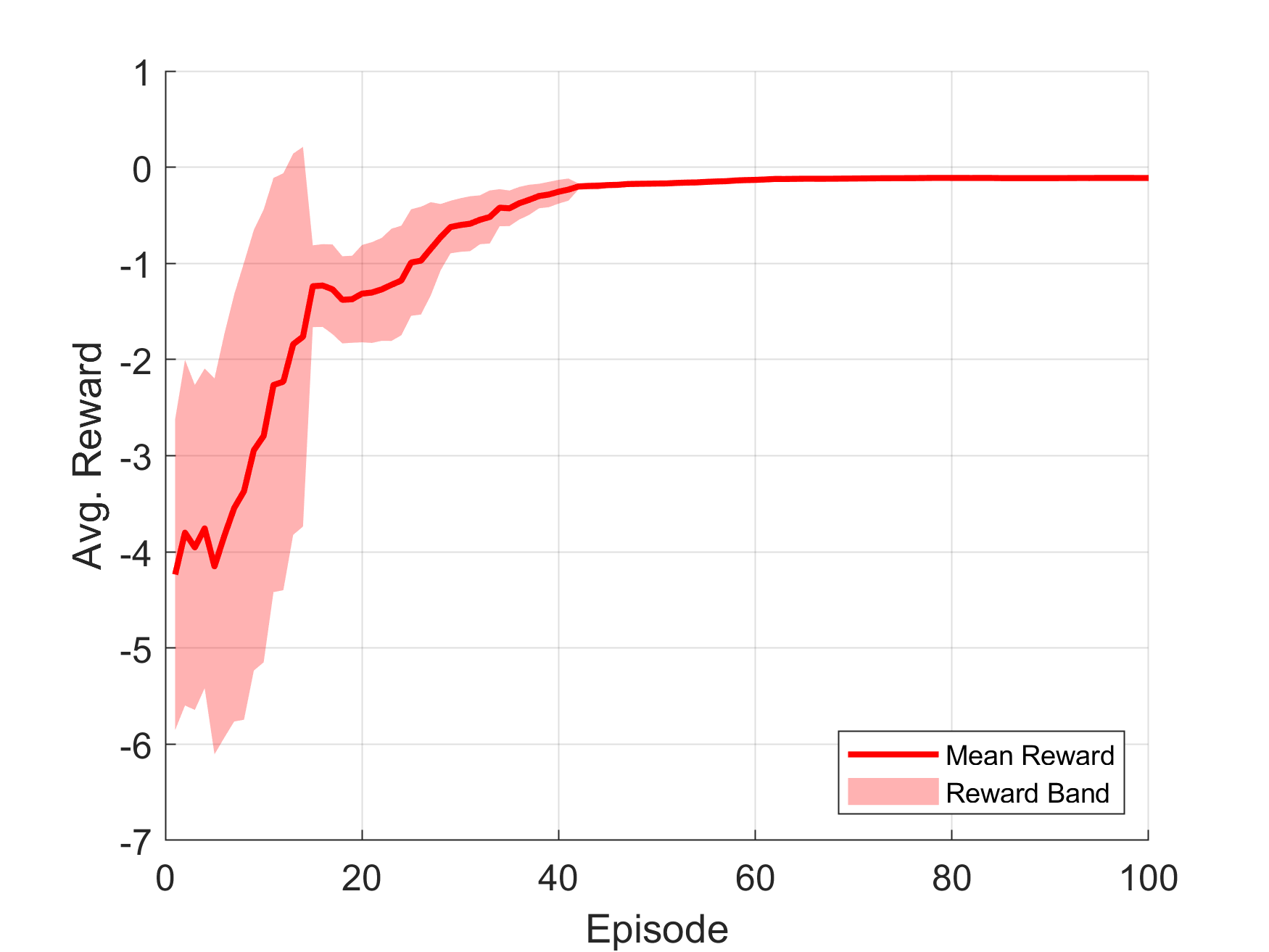} 
		\caption{Training reward of fixed-gain-DRL-model} 
		\label{fig:reward-fixed-gains}

	\end{minipage}\hfill
	\begin{minipage}{0.48\textwidth}
		\centering
		\includegraphics[width=1.0\textwidth]{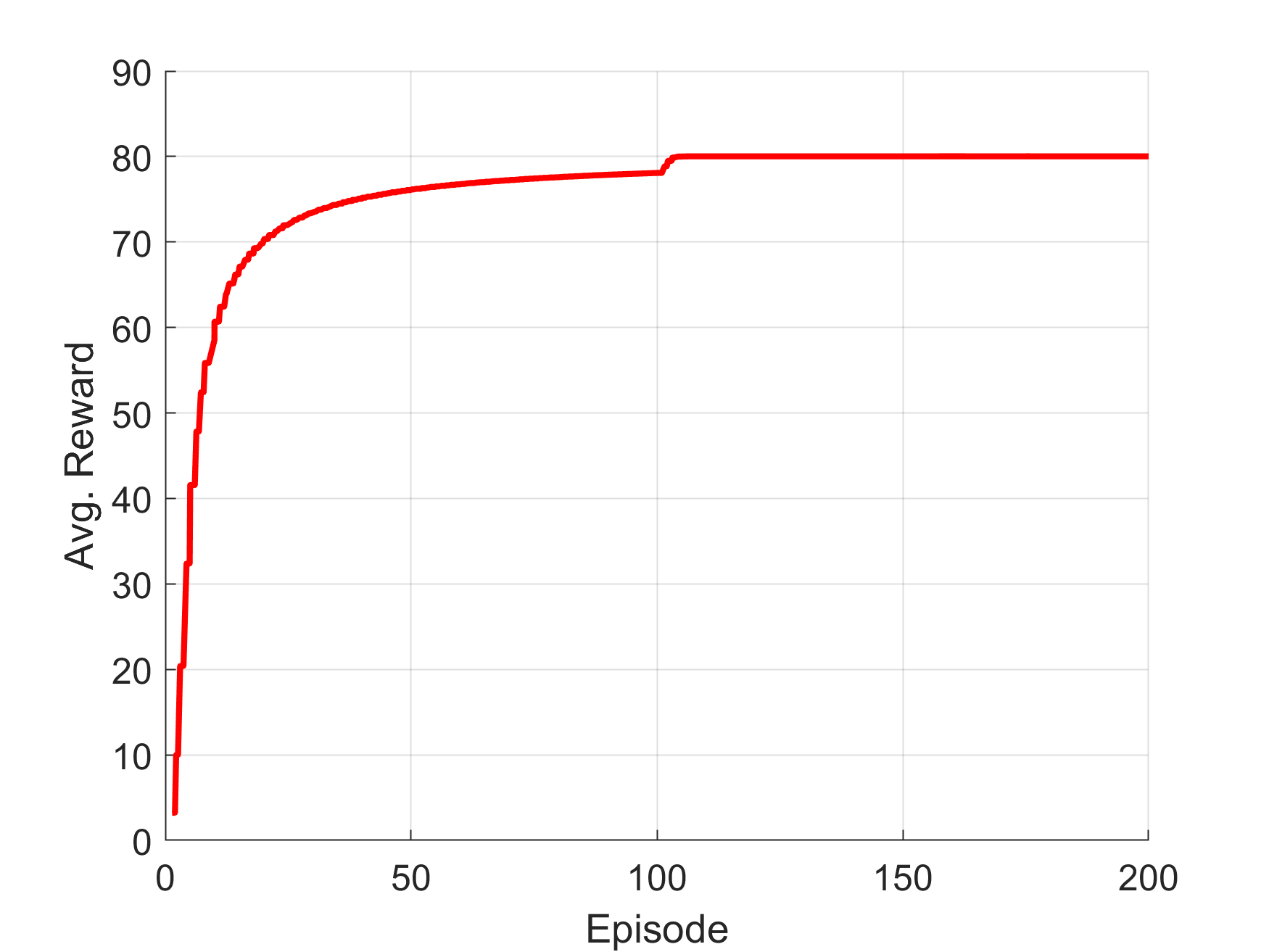}
		\caption{Training reward of adaptive-gain-DRL-model}
		\label{fig:adaptive-gains-rewards}
	\end{minipage}
\end{figure*}

A DRL agent based on the Proximal Policy Optimization (PPO) algorithm was trained using the actor network shown in Fig. \ref{fig:Current-PI-to-NN}. A fully connected neural network was employed as the critic network. After training, the model provided the grid-following (GFL) inverter's current controller gains \( K_p \) and \( K_i \) as the actor's policy weights. These generated weights were then transferred into the Simulink simulation to compare the fixed-gain-RL-agent's performance with Simulink EMT simulation. Finally, the current controller gains, \( K_p \) and \( K_i \) were deployed into the hardware testbed to verify the effectiveness of the optimal gains. 

Two sets of results were collected to compare the performance of the Simulink model using the optimal gains obtained from the RL agent's policy. The first set of results was gathered from the RL environment; these results are represented by the solid line in Fig. \ref{fig:fixed-gain-gfl-gfm-performance}. The second set was collected from the Simulink model, which used the grid-following (GFL) current controller gains \( K_p \) and \( K_i \) provided by the RL agent; these results are shown as dotted lines in the figure. The comparison clearly indicates that the generated fixed gains offer similar performance to the RL agent's current controller. Therefore, this result indicates that the optimum gains received from RL agent is directly deployable.  The fixed gains derived from the trained RL agent are presented in Table \ref{tab:RL-agent-fixed-gains}, which displays the initial gains and the optimal gains after training.

Fig. \ref{fig:reward-fixed-gains} illustrates the training performance of the fixed-gain RL agent used to tune the controller. The rewards increase with each episode, indicating that the model is consistently learning from the provided reward function. The figure shows both the reward bands and the mean reward over the episodes. The reward band indicates the fluctuations of the reward during training. After approximately 50 episodes, the rewards stabilize around -$0.1$. To evaluate the significance of this reward value, let's break down the reward function presented in Equation \ref{eqn:fixed-gain-reward}. The maximum possible reward is zero, which would occur if there were no difference between the reference setpoint and the measured dq-axis current, and if the actions (control inputs) were zero. However, in practice, the agent needs to generate dq-axis voltage to drive the inverter, and there is an initial settling time during which some error is inevitable. Therefore, achieving a zero reward is not feasible in a practical sense.

Given these considerations, a stabilized reward of -0.1 indicates that the RL agent has reached a near-optimal performance level. It suggests that the agent has effectively minimized the difference between the reference and measured currents and has learned to generate appropriate control actions, resulting in efficient tuning of the controller.

	

\begin{table}[tb!]
    \centering
    \caption{Trained GFL Current Controller Gains from RL Agent Actor Weights}
    \begin{tabular}{p{3cm} p{2cm} p{2cm}}
    \hline
    \textbf{Parameter Name} & \textbf{Initial Gains} & \textbf{Trained Gains} \\
    \hline
    $K_p$ & $1$ & $1.4406$ \\
    $K_i$ & $5$ & $12.7927$ \\
    \hline
    \end{tabular}
    \label{tab:RL-agent-fixed-gains}
\end{table}

\subsection{Adaptive-gain-DRL-model Simulation Performance}

\begin{figure*}[!ht]
	\centering
	\begin{minipage}{0.48\textwidth}
		\centering
		\includegraphics[width=1.0\textwidth]{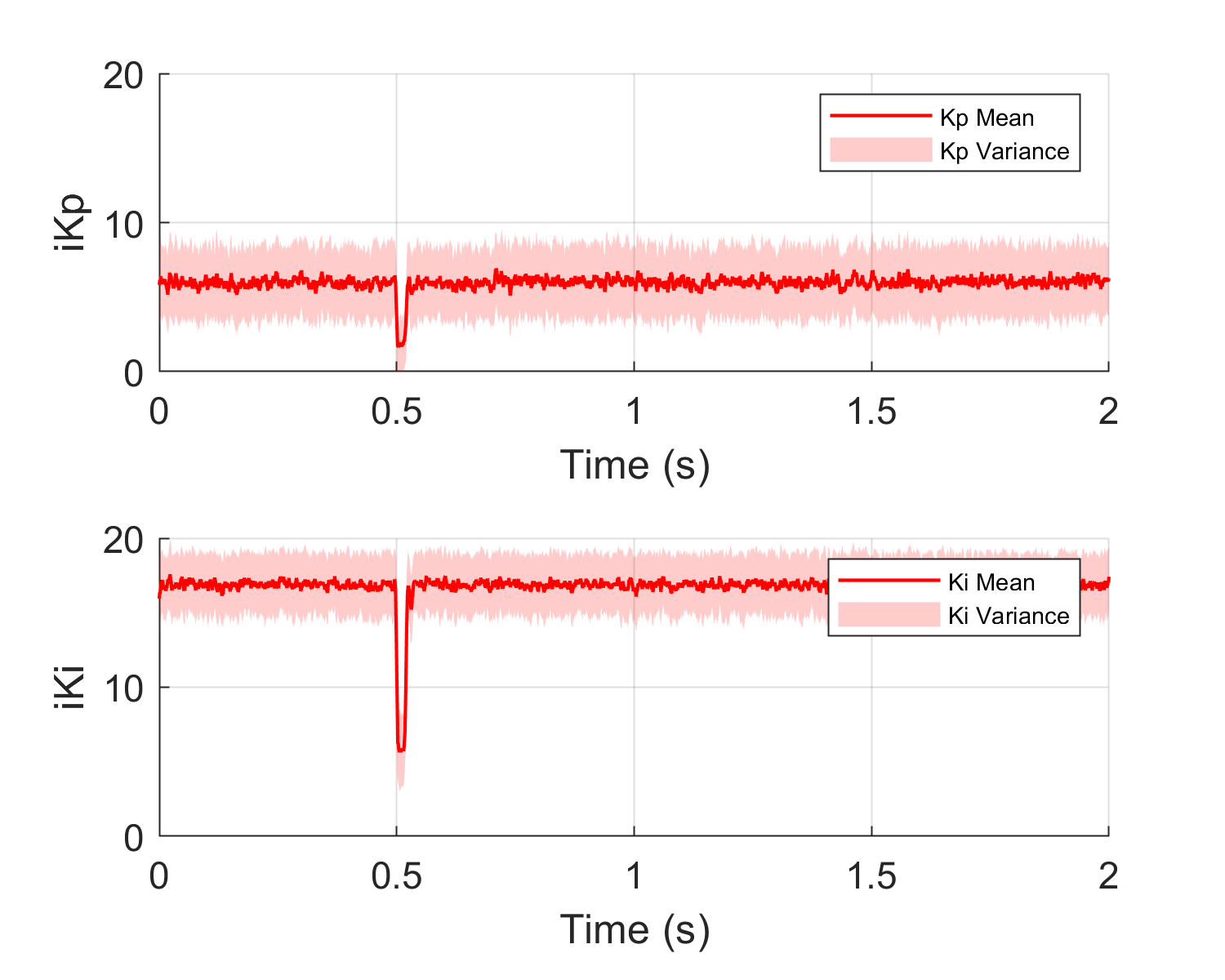}
		\caption{Adaptive gain generated as the RL agent action}
		\label{fig:adaptive-gain-kp-ki}
	\end{minipage}\hfill
	\begin{minipage}{0.48\textwidth}
		\centering 
		\includegraphics[width=1\textwidth]{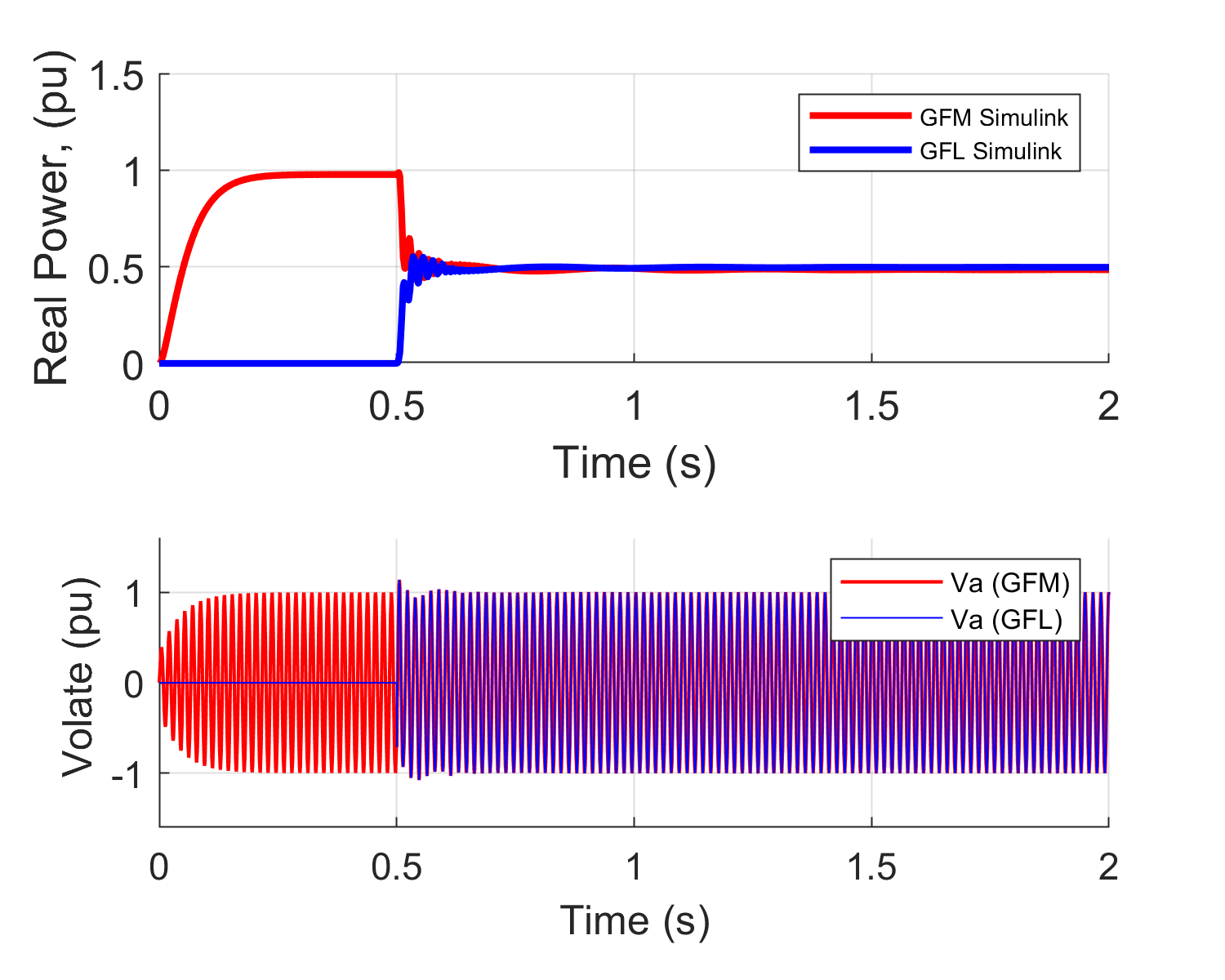} 
		\caption{Performance of the GFM and GFL utilizing a gain scheduling value utilized the adaptive gains $K_p, K_i$  mean values } 
		\label{fig:adaptive-gain-scheduling-performance} 
	\end{minipage}
\end{figure*}

Fig. \ref{fig:adaptive-gain-performance-python-single-run} presents the adaptive-gain-DRL-model performance when the model was trained and the agent was optimized. The results are collected from the Python environment utilizing the trained RL agent in a single run. It shows that the initial transient performance of the GFL converter is better then fixed-gain-DRL model performance. The transient overshoot of the GFL converter is reduced, and the settling time is significantly shortened.

The PPO RL agent provides adaptive gains as its action output. The RL agent is trained to generate gains based on the GFL inverter operating conditions. Fig. \ref{fig:adaptive-gains-rewards} illustrates the training performance of the RL agent used to tune the controller. The reward consistently increased with each episode, indicating that the model is effectively learning from the provided reward function. After approximately 100 episodes, the reward saturated indicates training converged.

The agent's actions were also recorded during the validation of the adaptive-gain-RL-agent, as depicted in Fig. \ref{fig:adaptive-gain-kp-ki}. The action variation is provided as a band. It was also found that the model's performance degrades if a sample of $K_p$ and $K_i$ is selected out of this band. Due to the page limitations this results are not presented. Therefore, this band indicates the GFL inverter's stability boundary for the $K_p$ and $K_i$ gains.

The adaptive gains generated by the RL agent as action outputs provided valuable insights. When the GFL inverter was paralleled at $t = 0.5$ s, the GFL current controller gains  $K_p$ and $K_i$ values reduced to less than half of the mean value. This indicates that the RL agent learned the system behavior and identified that the system performs better if the gains are reduced during transients. Therefore, to validate the performance in Simulink, the simplest approach is to deploy the RL agent's policy into the Simulink model. To simplify the process, the mean value of the agent's action is deployed into the Simulink model just like  a gain scheduling. Fig. \ref{fig:adaptive-gain-scheduling-performance} presents the performance of the GFL inverter utilizing the gain scheduling mechanism derived from the adaptive gain mean values presented in Fig. \ref{fig:adaptive-gain-kp-ki}. It is evident that the transient performance, voltage regulation, overshoot, and settling time all improved significantly. However, since the signal is hard-coded as a gain schedule, this method cannot mitigate transient situations during operation. Therefore, the agent's policy needs to be deployed directly into the model.

\subsection{Fixed-gain-DRL-model Hardware Verification}

\begin{figure} 
	\centering 
	\includegraphics[width=0.5\textwidth]{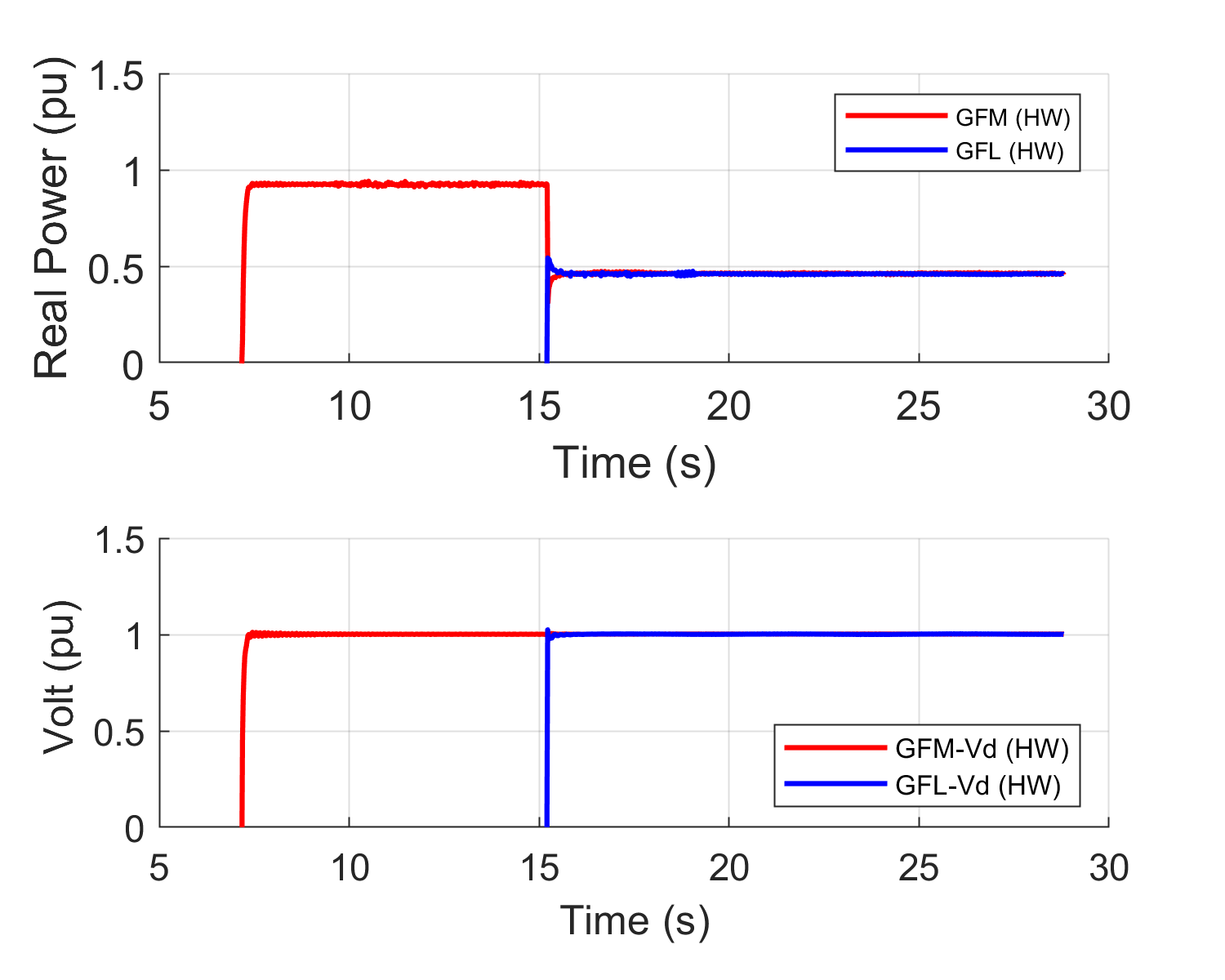} 
	\caption{Hardware performance of GFM and GFL using the fixed gains for the GFL current controller} 
	\label{fig:hw-fixed-gain} 
\end{figure}

Fig. \ref{fig:hw-fixed-gain} presents the hardware verification of GFM and GFL inverter performance using the fixed gains obtained from fixed-gain-RL-model. The current controller gains used in the GFL inverter are $K_p = 1.4406$ and $K_i = 12.7927$, as shown in Table \ref{tab:RL-agent-fixed-gains}. The GFM inverter was started first, forming the grid, and at \( t = 15s \), the GFL inverter was activated. It synchronized and began injecting power into the grid according to the setpoint. There was only a small spike at the beginning of the step response at \( t = 15s \), and the GFL inverter stabilized immediately. Interestingly, the hardware performance was even better than the simulation results shown in Fig. \ref{fig:fixed-gain-gfl-gfm-performance}.



\subsection{Adaptive-gain-DRL-model Hardware Verification}
The adaptive-gain-DRL-model provides a policy network after training which is a fully connected neural network responsible for generating real-time adaptive $K_p$ and $K_i$ values. For successful deployment, the network must compute at the same frequency as the control loop, which operates at 24 kHz. Due to the computational constraints of the microcontroller, further optimization of the network is required to ensure it meets the strict timing requirements. At present, the optimization process is ongoing to enable the network to fit within the microcontroller's limitations for hardware verification. 


\section{Discussion}
\label{discussion}

IBR power systems are inherently nonlinear, making it challenging for conventional control system analysis to determine control gains that perform optimally under all conditions. Traditional methods often fail to account for the complexities and the dynamic behaviors present in such systems. Therefore, innovative approaches are necessary to effectively tune IBR-based power systems.

Deep Reinforcement learning based controller tuning has gained popularity due to its ability to learn nonlinear behaviors. Neural networks within RL agents can adapt to various conditions, making them suitable for handling nonlinearity, fault conditions, and contingency scenarios. The advantage of this method is that the agent can be trained under different conditions, allowing it to manage complex situations that traditional controllers may not handle efficiently.



Several approaches have been reviewed in the literature for tuning control gains: (1) Training an RL agent to generate adaptive gains for the PI controller. This technique enables the controller to adjust its gains in real-time based on the operating conditions, improving performance during transients and disturbances. (2) Training an RL agent to produce the PI controller gains as neural network weights. This approach integrates the RL agent's learning directly into the controller's parameters, enabling seamless adjustments to system changes. (3) Training an RL agent to learn the PI control behaviors directly.


This research focuses on the first two methods, as they are the safest ways to operate power inverters. Although many literature \cite{xiongDeepReinforcementLearning2022, wooDSTATCOMDqAxis2021, schenkeControllerDesignElectrical2020, saadatmandAdaptiveCriticDesignbased2021, matlabTrainTD3Agent} explored using RL agents to replace traditional PI controllers in various inverter-based applications, we strongly critique this method because it is challenging to predict how the RL agent will behave in situations not encountered during training. Moreover, it is impractical to modify existing deployed high-power inverters that have already undergone extensive testing and certification processes.


As grid conditions evolve and the system loses inertia due to the retirement of conventional generators, situations will arise where existing inverters need retuning to suppress subsynchronous oscillations caused by weak grid conditions and control interactions. In such scenarios, the second method offers an excellent solution. It is feasible to model the entire grid with the inverter and then train the inverter gains using an RL agent, utilizing the fixed-gain method presented in this research.


Regarding the first method, while it may not be possible to integrate the adaptive gain policy directly with existing deployed inverters, it provides valuable insights into gain adjustments during faults and transients. This approach aids in selecting optimal gains capable of handling various situations effectively.

	
In this paper, we demonstrated how to tune the GFL current controller in a two-inverter setup. This work is part of a larger ongoing project focused on subsynchronous oscillation mitigation using reinforcement learning. In the future, we plan to implement this mechanism in a larger transmission grid to mitigate subsynchronous oscillations caused by control interactions.

\section{Conclusion}
\label{conclusion}


This research demonstrates a faster approach to tuning control parameters of inverter-based resource (IBR) power systems using reinforcement learning (RL) in a Python environment. The proposed methods provide an elegant solution that significantly reduces the RL agent's training time. By effectively handling the nonlinearities inherent in IBR power systems, the RL-based controller tuning addresses the limitations of conventional control system analysis, which often struggles to determine optimal control gains under all operating conditions. Applying this method to large-scale power systems can help mitigate subsynchronous control interactions caused by weak grid conditions, enhancing system stability and making the grid more resilient. Future work involves integrating the RL agent's policy into larger and more complex power system models. This integration would enable inverters to adapt dynamically to evolving grid conditions, further improving the resilience and reliability of power systems.

\section{Acknowledgment}
We extend our gratitude to the Electric Power Research Institute (EPRI) for their support in the hardware validation of this research.




\bibliographystyle{IEEEtran}
\bibliography{References.bib}

\end{document}